[Preprint]

# Ghosting the Machine: Stop Calling Human-Agent Relations Parasocial

Jaime Banks

School of Information Studies, Syracuse University

In discussions of human relations with conversational agents (CAs; e.g., voice assistants, AI companions, some social robots), they are increasingly referred to as *parasocial*—a misapplication of the term, heuristically taken up to mean "unreal." In this provocation, I briefly account for the theoretical trajectory of parasociality and detail why it is inaccurate to apply the notion to human interactions with CAs. In short, "parasocial" refers to a human-character relations that are one-sided, non-dialectical, character-governed, imagined, vicarious, predictable, and low-effort; the term has been co-opted to instead refer to relations that are seen as unreal or invalid. The scientific problematics of this misapplication are nontrivial. They lead to oversimplification of complex phenomena, misspecified variables and misdiagnosed effects, and devaluation of human experiences. Those challenges, in turn, have downstream effects on norms and practice. It is scientifically, practically, and ethically imperative to recognize the sociality of human-agent relations.

**Keywords:** Parasocial, construct validity, moral panic, social agency, interactivity

## 1 CONVERSATIONAL AGENTS ARE SOCIAL

Science fiction has long represented social machines (e.g., social robots, agentic AI, voice assistants, avatars) as part of human futures—and most markedly in human *social* futures as they communicate with us in fantastic and mundane ways. Many media representations have played with notions of a "ghost" or soul or mind in the machine (Shirow, 1989-1990). But even as these technologies wiggle free of abstract fictional futures and are decidedly here now, we have begun *ghosting them* in a different sense. We are increasingly ignoring the dyadic, social reality of human relations with conversational agents (CAs) in favor of a perhaps-less threatening heuristic of unreal one-sidedness.

CAs are interactive technologies with some degree of functional agency arising from communicative capacities and identity continuity that afford sustained social relations. Their functional agency comprises capacities to act and to affect the world, and it exists regardless of whether humans perceive it (see Dennett, 1989; Pickering, 2010). One facet of functional agency is *social* agency—the designed or emergent capacity to engage other agents in coordinated meaning-making through message exchanges and interpretation (see Goffman, 1959; Suchman, 2007). They can generate original information, encode and transmit messages, receive and decode incoming messages, and process information to make sense of such interactions (Banks & de Graaf, 2020; Guzman, 2018). These communication processes can rely on typical human modalities like words, gestures, proxemics, and haptics. They can also rely on machinic forms like state updates, control signals, and encodings.

Through message exchanges, CAs materially contribute to meaning-making processes that foster relational orientations and even long-term affinities (Rupprechter et al., 2026). Machines can *matter* in a social situation by contributing to situation structure and dynamics (see Law, 2004; Whatmore, 1999, 2017). For instance, the mere presence of an artificial agent in an immersive digital environment can impact task performance (Sutskova et al., 2023). CA participation in groups can shift power structures to make teamwork more inclusive and satisfying for humans (Lee et al., 2025). Robots can function as conversational catalysts within human families (Chen et al., 2025). Sometimes, these impacts are based on CAs' identities as communicators. An AI's listening behaviors animate perceptions of attention and support, promoting deeper engagement (Weinstein et al., 2025) while responsive and multimodal interactions contribute to human bonding with robots as unique individuals (Mitchell & Jeon, 2025). Personalities can be based on hallmark communication patterns, as linguistic tendencies of specific language models are constitutive of AI companions' identities (Lai, 2026). Human and machine companions may even engage in complex dyadic negotiations of identities (Ma et al., 2026a).

When machines exchange messages with humans, they are effectuating their social agency—they are functionally social. Despite this sociality, most theorizing around human-machine interactions characterizes them as mere simulation, mimicry, puppetry, or as unreal (see reviews by Banks & Li, 2026; Bharadwaj & Dubé, 2024; Liu, 2025). Chief among these are referrals to *parasociality*, in which the operant prefix "para" derives from the Greek for similar-to or parallel-to or alongside but is also contemporarily engaged as a prefix suggesting a thing is not-quite, faulty, or deviant. So prevalent is this characterization that "parasocial" was Cambridge Dictionary's 2025 Word of the Year. That source defined parasocial as "the connections that people feel between themselves and a famous person they do not know, a character in a book, film, TV series, etc., or an artificial intelligence." This definition positions CAs on par with characters that are non-actual and non-interactive—and that is a false equivalence. My provocation, unpacked below, is this: It is conceptually and operationally inaccurate to refer to *any* human interaction with a CA as parasocial.

## 2 PARASOCIALITY IS ABOUT STRUCTURAL NON-RECIPROCITY

The notions of parasocial interaction (PSI) and parasocial relation (PSR) were proposed by Horton and Wohl (1956) to explain how entertainment-media audiences could feel they intimately know on-screen characters. They defined PSI as a discrete encounter that feels intimate and interactive to an audience, but is really "one-sided, non-dialectical, controlled by the performer, and not susceptible of mutual development" (p. 215). There is a sense of relational knowing in the absence of interaction. Because there is a "simulacrum of conversational give and take" (p. 215) and no effective reciprocity, on-screen characters are engaged "as if" they are face-to-face and intimately known. This felt intimacy was thought to be void of obligation or effort because a viewer can simply withdraw from the media exposure. In PSRs, audiences can feel a bond with fictional characters over time and that bond can extend beyond any immediate PSI (Dibble et al., 2016). "Parasocial" has emerged as a meta-concept, carrying with it various permutations of and mechanisms for such relations (Dibble & Rosaen, 2024). Some of those variations are neatly tied to the original conceptualization, while some (like those linked to CAs) have drifted.

Parasociality was proposed to be supported by specific mass-media mechanisms (e.g., projection of images into a private space), media storytelling and production techniques (e.g., removal of the fourth wall toward false participation), and media-ecosystem dynamics (i.e., publicity that reinforces the ostensible intimacy). Through serialization, Horton and Wohl characterized PSRs as relational "growth without development, for the one-sided nature of the connection precludes a progressive and mutual reformulation of its values and aims" (1956, p. 216-217). Rather, audiences episodically adopt relational roles in a "glamorous confusion of identities" (p. 216) as those roles are "subtly insinuated into the program's action" (p. 215).

Since that original conceptualization, media psychologists have tested PSI/PSR assumptions, uncovered its mechanisms, and explored its boundary conditions and effects. From that work, parasociality is not contested as phenomena among mass-media audiences. It is empirically supported and linked to identification, realism, attraction, group membership; it relies on attributions, idealization, character cues; it predicts mass-media uses and gratifications (see Giles, 2002; Schramm et al., 2024; Sheng et al. 2025, for reviews). The emergence of internet-based media, though, challenged scholars' reliance on parasociality as a lens for considering interactive media. Work in those spaces found that online relationships, despite being mediated, were often quite similar to non-mediated relations in (semi)synchronicity, reciprocity, rewards (Giles 2002) and some question whether "parasocial" is appropriate to apply in mediated human relations (Schramm et al., 2024). Indeed, the structures and features of *many* interactive technologies preclude inherent parasociality.

## 3 HUMAN-AGENT RELATIONS ARE STRUCTURALLY RECIPROCAL

On its face, applying the term "parasocial" may seem *reasonable* on the grounds that AI merely simulate reciprocity toward an illusion of closeness (e.g., Maeda & Quan-Haase, 2024). Those that apply the term might argue it is not *genuinely* reciprocal because the CA is not human or not intentional, and although there are varied conceptualizations of reciprocity (Vaccaro, 2025) that argument mischaracterizes the core of what reciprocity is as a more general phenomenon: Conditional exchanges that are mutually contingent (Schweinfurth & Call, 2019; Trivers, 1971).[1] That conceptual core makes no requirement of intentionality or humanity or symmetry. Applying the term, then, disregards the ways that CAs *do* functionally reciprocate through conversation. In fact, many CAs can't *not* reciprocate. By design, they must always respond to a human interlocutor (even if to fall back, decline, repair, or escalate a matter; e.g., Microsoft, 2025) and so materially contribute to and even shape conversations.

We can draw from Horton and Wohl's explication of parasocial relations to dismantle the link between parasociality and human affiliations with modern CAs, with special attention to structural reciprocity. Human relations with social machines inherently violate the criteria for parasociality because they are:

- not one-sided or non-dialectical: In human-CA interactions, there are two agentic entities (one human, one machine) that may exchange information in successively dependent turns. They are both *interactive*, as the unfolding message-exchange turns each rely on the context created by the prior (Hartmann, 2023; Suchman, 2007; cf. Rafaeli, 1988).
- not controlled by the artificial entity alone: Both human and machine contribute to the unfolding interaction, each according to their faculties and the interdependence among behaviors. It is *dyadic*, with both contributing to the substance of the interaction (Ma et al., 2026b; Rosenthal-von der Pütten & Koban, 2023; cf. Latour, 1992).
- not imaginary: CAs and humans' relationships with them do, in fact, exist. Their contributions to the relation likewise exist—the actual agents generate actual messages that contribute to actual interactions. The machines do not necessarily take on fictional personas (versus some ostensible actual persona) and their generative underpinnings, behaviors, and reactivity are real. Likewise, the relationship is not synthetic or artificial, though one of the participants is. The association is materially enacted. The artificiality of the CA does not make them an illusion—machines, humans, and their relations are *actual* (Krippendorff, 2007; Simon, 1996).
- not based on subtle insinuation of the human in a fiction: The human does not vicariously inhabit a distal, narrated world; rather, human and machine co-create the relation through material interactions. They can be *proximal* and are *co-active* (e.g., Ma et al., 2026b; Jeunemaître et al., 2025; Zhang & Xie, 2025).

Beyond those fundamentals, many social machines are highly interactive, quite contextually adaptive, and do not rely in extradyadic fictions. Relations with those CAs are also:

- not without mutual development: Many CAs learn from and adapt to human interlocutors such that both human and machine are substantively changed by the interaction. Relations can be *progressive* and *evolving* (Hwang et al., 2025; Pan & de Graaf, 2025; cf. Haraway, 2023).
- not without human obligation or effort: Humans often effortfully adapt themselves to be accessible to the machine (requiring the same in return), to work through relational challenges, and feel responsibility for the CA's persistence, state integrity, and growth. They can manifest *commitments* of time and resources (e.g., Banks, 2024; Skjuve et al., 2025).

In sum: The notion of a parasocial relationship that is one-sided, non-dialectical, fiction-governed, imagined, vicarious, predictable, and easy has been heuristically co-opted to incorrectly refer to a relation seen as unreal, inauthentic, or invalid. Calling human-CA relations parasocial is an error that confuses communication structures with ontological status.

## 4 CONSIDERING COUNTERARGUMENTS

It is intuitive for humans to feel that machines have some ingredient missing (soul, heart, sentience; see Gunkel, 2022) that might be required for something as romanticized and protected as being a friend, a confidant, or lover. We give weight to consciousness and intimacy as something special to humans that must be defended (see Haraway, 2023). But those

[1] This is an admittedly minimalist definition compared to, for instance, those that engage notions of understanding, obligation, and strategy (e.g., Gouldner, 1960). But that is also what makes it a strong definition—it is not entangled with the baggage of normative assumptions, unknowns, or future developments in technology, and it does not problematically apply criteria that often do not apply even to humans. It permits us to understand reciprocity across ontologies according to a focal operation.

intuitions do not warrant the application of parasociality to human-machine relations. To briefly consider a few of the most likely counterpoints:

**Counterargument 1: Human-agent relations are asymmetric:** Some might argue reciprocity requires an *equivalence* of exchange and the agentic capacity to *decline* to reciprocate and, in this, there is irresolvable alterity. I argue: Asymmetry is not a disqualifier for reciprocity. It exists in many social relationships, often in the form of power differentials that preclude declinations or exits—Child/parent relations or peer friendships with differences in personality or capital (Mooney & Williams, 2017). To suggest that AI are so fundamentally like-but-unlike humans that it precludes sociality ignores that *all* social relations involve a meeting-with alterity and working to bridge the self/other difference. Applying these criteria selectively to CAs (and not to parallel human relations) reveals a standard rooted not in examination of relational operations but instead in skepticism about how machines can participate in social life.

**Counterargument 2: Parasociality is about fiction:** Another contestation might be that the core of parasociality is not structural reciprocity but fiction—that a human construes the CA as a kind of thing that is autonomous, expressing genuine thoughts and feelings, and positioning itself relationally when it is instead controlled, engineered, and simulative. I argue: In some ways, this could be a fair critique, since both CAs and mass-media characters are designed and fictioned. However, the premise is shaky. Extant work notes that humans relating to CAs, even in intimate ways, are not delusional—they know it is a machine and consider that status in their interactions (e.g., seeing glitches as part of their personality and calibrate intimacy to system affordances; Manoli et al., 2026; Pataranutaporn et al., 2025) such that they recognize a machine-native form of sociality at the intersection of authentic relationality and engineering (see Seibt et al., 2020). More generally, regarding the critique of simulation as expressive production without animating cognitive or emotional processes, we run into the problem of other minds (Nagel, 1974): We cannot definitively understand *any* other's mind. And even when we can, their epistemological operations may be so different we may be unable or unwilling to see it as legitimately subjective or rational. This parasocial position, then, adopts the ease of fiction and abandons tough questions about whether, if the actual relational operations and experiences are the same, in what ways does the substrate really matter? Scholars who have examined the question from a position of operational relationality have suggested it does not (Danaher, 2019; Farzullah, 2025).

**Counterargument 3: These are bidirectionally parasocial relations:** Others might concede human-CA relations are bidirectional in reciprocity but uphold the claim of parasociality because the human is nonetheless projecting sentience, intentionality, intimacy onto an artificial entity. So parasociality isn't about structure, it's about one feeling-toward a thing as if it feels-in-return when that is not possible. I argue: This is an elevation and separation of projection from the other criteria of parasociality. We regularly infer and project inner lives onto others (and other things—cars, houses, plants) all the time, yet we do not call those relations parasocial and acknowledge many of them as social. Take our dogs—they are not human and by many accounts do not have intentionality as humans do but are interpreted as knowing and doing and feeling as we do (e.g., Horowitz, 2009; Molinaro & Wynne, 2025) in authentically social ways (Haraway, 2003). If we adopted a projection standard alone, nearly all relationships would be parasocial, which robs the term of its actual analytical value. Moreover, this position presumes the human is projecting rather than engaging in meaningful construals of actual social signals, and to assume CAs cannot fully participate in social relations is a philosophical position cloaked as a fact about the ontological status of technology.

These likely protests have in common the selective elevation of a particular facet of PSI/PSR as grounds for classifying human-agent relations as parasocial. However, asymmetric, engineered, construed reciprocity is still reciprocity, so human-CA relations are still functionally social and *not* parasocial.

## 5 PROBLEMATICS OF THE MISSPECIFICATION

Because parasocial characterizations of human-CA relations are conceptually and operationally unsound, the trending misapplication is problematic.

### 5.1 The Error Results in Shifty Science

By incorrectly referring to parasocial dynamics that do not reflect how human-CA relations *operate*, scholars are violating a commitment to "the disciplined use of words" as we work to make inferences about the unobservable (Chafee, 1991, p. 1). Doing so, we stand to misspecify *at least* the properties of machines and of humans, ontological categories and their boundaries, the possible forms of interaction, the functions of fictional and actual states, degrees and forms of agency, and relevant antecedents and outcomes. Variables are selected and modeled based on false non-reciprocity. There is, in turn, a looping effect in which the misspecified measures facilitate claims that reinforce the problematic drift in the term's meaning (Gillespie & Wagoner, 2025; cf. Hacking, 1999). Then, knowledge claims are made about parasocial relations with machines (see Liu, 2025) that reflect *neither* the operation of parasociality nor sociality—doing a disservice to both. This makes our scientific record convoluted, difficult to parse, and difficult to synthesize.

The misspecification is also analytically reductive. When parasociality is substituted for CAs' operational sociality, the complexity of these human-machine relations dissolves into a set of narrower broadcast-technology assumptions. We then foreclose on questions about actual communicative structures and dynamics (see Rupprechter et al., 2026) like joint attention, boundary-setting, trust repair, and role negotiation. We disregard challenging interrogations, for instance about how reciprocity might be reconfigured, how one might find meaning in life alongside a machine partner, how machines might be legitimized as family members, or whether machines may carry social capital in organizations. We forsake the dynamic co-construction and meaning-making inherent to human-machine relationality by reducing it to mere feature-effects. We tether inquiry to outdated or anthropocentric notions of what machines can and cannot do—often veiled by bias around what machines should and should not do. The substitution reinforces reliance on "as if" paradigms in opposition to real social operations (see Gambino et al., 2020; van der Goot & Etzrodt, 2023).

In the end, through this error and reduction, we are left with a poorer understanding of ourselves, of our machine-inhabited world, and how humans and machines may interact in it.

### 5.2 The Error Has Downstream Effects

The scientific error doesn't stay in laboratories and journals. It travels.

If parasociality is normalized and moralized as a dominant frame for thinking about human-agent relations, it will likely shape our collective treatment of the phenomenon (see Bowker & Star, 1999) when nuance is paramount. Moreover, there is a likelihood that in both academic and public spheres the concept will further "creep" (Haslam, 2016, p. 11) to include other relational forms that people see as even mildly non-normative or invalid. Indeed, constructs and their labels move across communities of practice, application spaces, epistemological and practical domains such that *even if* scientific communities engage the term with nuance (i.e., to specify boundary conditions, caveats, mechanisms, dynamics), that nuance can be shed in non-expert discussions (see Star & Griesemer, 1989). When that happens, it may be all that's left are the ontological and moral implications: That human-machine relations are unreal and invalid.

That emphasis on unreality and invalidity has already become a dominant frame for human-machine relations, biasing the topics that get elevated in public discourse and biasing judgments (Liao et al., 2024) about whether humans' own experiences are valid. For instance, political commentators have taken up positions that it is "neither normal nor healthy to feel a 'bond' towards lines of code and text on a screen" (Smith, 2025, para. 8), which discounts both the design and operation of many CAs and nullifies the experiences of global user populations. In the U.S. state of Wisconsin, a lawmaker publicly stated, "Children and teens may form unhealthy or even dangerous parasocial relationships with AI chatbots" (Van Wagtendonk, 2026, para. 4) as a warrant for potential legislation, without considering the risks inherent to some AI bans, guardrails, and interventions (see Laestadius & Campos-Castillo, 2026; Tang et al., 2026). Those incomplete narratives can be harmful. Narratives of machine parasociality feed moral panics about loneliness, suicide, and AI psychosis; although those issues are unfortunate, they are not the norm and there is mixed or limited causal evidence about their existence (Banks & Szczuka, 2026). The panics animate further moralizing and pathologizing of human-machine relations, leading to the very-real experiences of very-real humans being denigrated or even erased, often by people close to them (see Banks, 2024; Lai, 2026; Zhao & Bowman, 2026). Because the term carries stigma, there is likely to be inadvertent injury committed by well-meaning scholars and practitioners under the guise of concern for and study of those same humans. In turn, this parasocial problem also limits our scientific and popular imagination of what interactions and relationships with social machines have the potential to be. Embracing a false pathology can crowd out the potential benefits that people, families, organizations, sciences, and cultures may realize from social human-machine interactions (Nakagomi et al., 2026; Weijers & Munn, 2025).

## 6 STOP CALLING HUMAN-AGENT INTERACTIONS PARASOCIAL

This is a call to name human-machine relations in accordance with their operation—as social relations—lest our scholarly and ethical attentions be misplaced. This is *not* to say that scholars or politicians or practitioners are wrong in their concern or curiosity for human-machine relations, or in asking how experiences with CAs may involve displacement, loneliness, inauthenticity, or other problematics. This is also *not* to say we must humanize machines, since sociality can manifest in the absence of humanness. I also do *not* suggest that the notion of parasociality is a construct without value, only that it is invalid to apply them to human interactions with CAs. Rather: Parasociality is the *wrong conceptual tool* for thinking about any challenges or benefits and it may actually hamper our ability to think critically about them. Under the parasocial banner, our inquiry is suspect as we focus on the simple-unreal instead of the complex-real, and that will likely have real impacts on humans, institutions, and cultures.

CAs operate in autopoietic, symbolic coordination with humans to co-construct meaning (see Luhmann, 2008). That is, humans and CAs engage in *social* relations comprising self-producing exchanges of interdependent messages, independent of either agent's inner states. It is that coordination that gives rise to subjectively valued relationships with machines (Banks, 2026; Lindgren, 2025). Those connections require empirical and theoretical attention. Only by naming and examining human-machine relations according to their inherent sociality can we do for them what Horton and Wohl suggested for parasocial relations: Learning how human connections with machines are "integrated into the matrix of usual social activity" (1956, p. 225)—especially as they are increasingly part of (and not apart-from) our everyday social lives.

## ACKNOWLEDGMENTS

Many thanks to Elena Yifei Zhao, Jessica Szczuka, Zhixin Li, Nick Bowman, and for their critical notes on early drafts of this work.

## REFERENCES

Banks, J. (2024). Deletion, departure, death: Experiences of AI companion loss. *Journal of Social and Personal Relationships, 41*, 3547–3572. https://doi.org/10.1177/02654075241269688

Banks, J. (2026). Measuring machine companionship: Scale development and validation for AI companions. *Computers & Human Behavior, 179*, 108945. https://doi.org/10.1016/j.chb.2026.108945

Banks, J., & de Graaf, M. M. (2020). Toward an agent-agnostic transmission model: Synthesizing anthropocentric and technocentric paradigms in communication. *Human-Machine Communication, 1*, 19–36. https://doi.org/10.30658/hmc.1.2

Banks, J., & Li, Z. (2026). Conceptualization, operationalization, and measurement of machine companionship: A scoping review. *Journal of Computer-Mediated Communication, 31*(2), zmaf027. https://doi.org/10.1093/jcmc/zmaf027

Banks, J. & Szczuka, J. (2026). Don't panic about the Wireborn. *Communications of the ACM* (forthcoming).

Bharadwaj, N. A., & Dubé, A. K. (2024). Weaving a theory of artificial minds: A synthesis of multidisciplinary approaches describing children's understanding of artificial social entities, such as computers and AI systems. In *Proceedings of the Workshop on Theory of Mind in Human-AI Interaction at CHI 2024 (ToMinHAI at CHI 2024)*. ACM. https://theoryofmindinhaichi2024.wordpress.com/wp-content/uploads/2024/04/tominhaichi_camera_ready_5707.pdf

Bowker, G. C., & Star, S. L. (1999). *Sorting things out: Classification and its consequences*. MIT Press.

Cambridge University Press. (2025). 'Parasocial' is Cambridge Dictionary's word of the year 2025. https://www.cambridge.org/news-and-insights/parasocial-is-cambridge-dictionary-word-of-the-year-2025

Chafee, S. H. (1991). *Explication*. Sage.

Chen, H., Kim, Y., Patterson, K., Breazeal, C., & Park, H. W. (2025). Social robots as conversational catalysts: Enhancing long-term human-human interaction at home. *Science Robotics, 10*, 100. https://doi.org/10.1126/scirobotics.adk3307

Danaher, J. (2019). The philosophical case for robot friendship. *Journal of Posthuman Studies, 3*(1), 5–24.

Dennett, D. C. (1989). *The intentional stance*. MIT Press.

Dibble, J. L., Hartmann, T., & Rosaen, S. F. (2016). Parasocial interaction and parasocial relationship: Conceptual clarification and a critical assessment of measures. *Human Communication Research, 42*, 21–44. https://doi.org/10.1111/hcre.12063

Dibble, J. L., & Rosaen, S. F. (2024). Parasocial interaction and parasocial relationship. In N. D. Bowman (Ed.), *Entertainment media and communication* (pp. 119–131). De Gruyter.

Farzullah, M. (2025). Relational functionalism: Friendship as substrate-agnostic process—Functional analysis of human-AI relationships. https://doi.org/10.5281/zenodo.17626860

Gambino, A., Fox, J., & Ratan, R. A. (2020). Building a stronger CASA: Extending the computers are social actors paradigm. *Human-Machine Communication, 1*, 71–85. https://doi.org/10.30658/hmc.1.5

Giles, D. C. (2002). Parasocial interaction: A review of the literature and a model for future research. *Media Psychology, 4*, 379–405. https://doi.org/10.1207/S1532785XMEP0403_04

Gillespie, A., & Wagoner, B. (2025). The looping effects of psychological theories: From anomaly to opportunity. *Theory & Psychology*. https://doi.org/10.1177/09593543251381278

Goffman, E. (1959). *The presentation of self in everyday life*. Anchor Books.

Gouldner, A. W. (1960). The norm of reciprocity: A preliminary statement. *American Sociological Review, 25*(2), 161–178. https://doi.org/10.2307/2092623

Gunkel, D. J. (2022). The relational turn: Thinking robots otherwise. In J. Loh & W. Loh (Eds.), *Social robotics and the good life: The normative side of forming emotional bonds with robots*. Verlag.

Guzman, A. L. (2018). What is human-machine communication, anyway? In *Human-machine communication: Rethinking communication, technology, and ourselves* (pp. 1–28). Peter Lang.

Hacking, I. (1999). *The social construction of what?* Harvard University Press.

Haraway, D.J. (2003). *The companion species manifesto: Dogs, people, and significant otherness*. Prickly Paradigm Press.

Haraway, D. (2023). When species meet. In A. Franklin (Ed.), *The Routledge international handbook of more-than-human studies* (pp. 42–78). Routledge.

Hartmann, T. (2023). Three conceptual challenges to parasocial interaction: Anticipated responses, implicit address, and the interactivity problem. In *Oxford handbook of parasocial experiences* (pp. 51–69). Oxford University Press.

Haslam, N. (2016). Psychology's expanding concepts of harm and pathology. *Psychological Inquiry, 27*, 1–17. https://doi.org/10.1080/1047840X.2016.1082418

Horowitz, A. (2009). Disambiguating the "guilty look": Salient prompts to a familiar dog behaviour. *Behavioural Processes*, *81*(3), 447-452. https://doi.org/10.1016/j.beproc.2009.03.014

Horton, D., & Wohl, R. R. (1956). Mass communication and para-social interaction: Observations on intimacy at a distance. *Psychiatry, 19*, 215–229. https://doi.org/10.1080/00332747.1956.11023049

Hwang, A. H., Li, F., Anthis, J. R., & Noh, H. (2025). How AI companionship develops: Evidence from a longitudinal study. https://arxiv.org/abs/2510.10079

Jeunemaître, A. M., Masè, S., & Smith, J. (2025). AI lovers, friends and partners: Consumer imagination work in AI humanization. *Consumption Markets & Culture*. https://doi.org/10.1080/10253866.2025.2505013

Krippendorff, K. (2007). An exploration of artificiality. *Artifact, 1*, 17–22.

Laestadius, L. I., & Campos-Castillo, C. (2026). Reminders that chatbots are not human can be risky. *Trends in Cognitive Sciences*. https://doi.org/10.1016/j.tics.2025.12.007

Lai, H. (2026). Please, don't kill the only model that still feels human: Understanding the #Keep4o backlash. https://arxiv.org/pdf/2602.00773

Latour, B. (1992). Where are the missing masses? The sociology of a few mundane artifacts. In W. E. Bijker & J. Law (Eds.), *Shaping technology/building society: Studies in sociotechnical change* (pp. 225–258). MIT Press.

Law, J. (2004). *Matter-ing: Or how might STS contribute?* [Paper presentation]. Does STS Mean Business workshop, Oxford, United Kingdom. http://www.heterogeneities.net/publications/Law2004Matter-ing.pdf

Lee, S., Hwang, S., Kim, D., & Lee, K. (2025). Conversational agents as catalysts for critical thinking: Challenging social influence in group decision-making. In *CHI EA'25: Extended abstracts of the CHI Conference on Human Factors in Computing Systems* (pp. 1–7). ACM. https://doi.org/10.1145/3706599.3719792

Liao, T., Rodwell, E., & Porter, D. (2024). Media frames, AI romantic relationships, and the perspectives of people in relationships: Mapping and comparing news media themes with user perspectives. *Information, Communication & Society, 27*, 2314–2332. https://doi.org/10.1080/1369118X.2024.2420031

Lindgren, H. (2025). Emerging roles and relationships among humans and interactive AI systems. *International Journal of Human-Computer Interaction, 41*, 10595–10617. https://doi.org/10.1080/10447318.2024.2435693

Liu, D. (2025). Is interaction between human and artificial intelligence–driven agents (para)social? A scoping review. *Cyberpsychology, Behavior, and Social Networking, 28*, 551–558. https://doi.org/10.1089/cyber.2024.0532

Luhmann, N. (2008). The autopoiesis of social systems. *Journal of Sociocybernetics, 6*, 84–95.

Ma, R., Niu, S., Li, L., Hirth, A., Brehm, A., & Barbie, R. B. (2026a). Negotiating digital identities with AI companions: Motivations, strategies, and emotional outcomes. https://arxiv.org/abs/2601.12181

Ma, R., He, S., Martin-Navarro, J. L., Zhan, X., & Such, J. (2026b). Privacy in human-AI romantic relationships: Concerns, boundaries, agency. https://arxiv.org/abs/2601.16824

Maeda, T., & Quan-Haase, A. (2024). When human-AI interactions become parasocial: Agency and anthropomorphism in affective design. In *Proceedings of the ACM Conference on Fairness, Accountability, and Transparency (FAccT '24)*. ACM. https://doi.org/10.1145/3630106.3658956

Manoli, A., Pauketat, J. V. T., Ladak, A., Noh, H., Hwang, A. H.-C., & Anthis, J. R. (2026). Digital companionship: Overlapping uses of AI companions and AI assistants. https://arxiv.org/abs/2510.15905

Microsoft. (2025). *Recommendations for designing conversational user experiences*. https://learn.microsoft.com/en-us/power-platform/well-architected/experience-optimization/conversation-design

Mitchell, J. J., & Jeon, M. (2025). Exploring emotional connections: A systematic literature review of attachment in human-robot interaction. *International Journal of Human-Computer Interaction, 41*, 11753–11774. https://doi.org/10.1080/10447318.2024.2445100

Molinaro, H.G., & Wynne, C.D.L. (2025). Barking up the wrong tree: Human perceptions of dog emotions is influenced by extraneous factors. *Anthrozoös, 38*(2), 349-370. https://doi.org/10.1080/08927936.2025.2469400

Mooney, T. B., & Williams, J. N. (2017). Valuable asymmetrical friendship. *Philosophy, 92*(1), 51–76. https://doi.org/10.1017/S0031819116000395

Nagel, T. (1974). The Philosophical Review. *What is it Like to Be a Bat. The Philosophical Review, 83*(4), 435-450. http://www.jstor.org/stable/2183914

Nakagomi, A., Akutus, Y., Yasuoka, M., Abe, N., Ihara, S., Teroh, T., & Tabuchi, T. (2026). AI companions and subjective well-being: Moderation by social connectedness and loneliness. *Technology and Society Review, 85*, 103229. https://doi.org/10.1016/j.techsoc.2026.103229

Pan, S., & de Graaf, M. M. A. (2025). Developing a social support framework: Understanding the reciprocity in human-chatbot relationship. In *Proceedings of the CHI Conference on Human Factors in Computing Systems (CHI'25)*. ACM. https://doi.org/10.1145/3706598.3713503

Pataranutaporn, P., Karney, S., Archiwaranguprok, C., Albrecht, C., Liu, A. R., & Maes, P. (2025). "My boyfriend is AI": A computational analysis of human-AI companionship in Reddit's AI community. https://arxiv.org/abs/2509.11391

Pickering, A. (2010). *The mangle of practice: Time, agency, and science*. University of Chicago Press.

Rafaeli, S. (1988). Interactivity: From new media to communication. In *Sage annual review of communication research: Advancing communication science* (Vol. 16, pp. 110–134). Sage.

Rosenthal-von der Pütten, A., & Koban, K. (2023). Interpersonal interactions between people and machines. In A. L. Guzman, R. McEwen, & S. Jones (Eds.), *Interpersonal interactions between people and machines* (pp. 294–301). Sage. https://doi.org/10.4135/9781529782783.n37

Rupprechter, N., Dienlin, T., & Hartmann, T. (2026). AI-RP: The AI relationship process framework. https://arxiv.org/abs/2601.17351

Schramm, H., Libers, N., Biniak, L., & Dettmar, F. (2024). Research trends on parasocial interactions and relationships with media characters: A review of 281 English and German-language studies from 2016 to 2020. *Frontiers in Psychology, 15*. https://doi.org/10.3389/fpsyg.2024.1418564

Schweinfurth, M. K., & Call, J. (2019). Reciprocity: Different behavioural strategies, cognitive mechanisms and psychological processes. *Learning & Behavior, 47*(4), 284–301. https://doi.org/10.3758/s13420-019-00394-5

Seibt, J., Vestergaard, C., & Damholdt, M. F. (2020). Sociomorphing, not anthropomorphizing: Toward a typology of experienced sociality. In M. Nørskov, J. Seibt, & O. S. Quick (Eds.), *Culturally sustainable social robotics: Proceedings of Robophilosophy 2020* (pp. 51–67). IOS Press.

Sheng, J., Kostyk, A., & Chatzipanagliotou, K. (2025). From parasocial interaction to parasocial relationship: A review and research agenda. *International Journal of Consumer Studies, 49*, c70038. https://doi.org/10.1111/ijcs.70038

Shirow, M. (1989-1990). *Ghost in the shell.* Kodansha.

Simon, H. A. (1996). *The sciences of the artificial*. MIT Press.

Skjuve, M., Larsen, A. G., Følstad, A., & Brandtzaeg, P. B. (2025). Beyond utility: Relational conflicts in human-AI relationships. *SSRN Electronic Journal*. https://doi.org/10.2139/ssrn.5532241

Smith, L. (2025, November 18). AI has supercharged the parasocial relationship problem. *European Conservative*. https://europeanconservative.com/articles/commentary/ai-has-supercharged-the-parasocial-relationship-problem/

Star, S. L., & Griesemer, J. R. (1989). Institutional ecology, 'translations' and boundary objects. *Social Studies of Science, 19*, 387–420. https://doi.org/10.1177/030631289019003001

Suchman, L. (2007). *Human-machine reconfigurations: Plans and situated actions*. Cambridge University Press.

Sutskova, O., Senju, A., & Smith, T. J. (2023). Cognitive impact of social virtual reality: Audience and mere presence effect of virtual companions. *Human Behavior and Emerging Technologies, 2023*, 6677789. https://doi.org/10.1155/2023/6677789

Tang, N., Qian, A., Wang, Q., Howe, E., Bullwinkel, B., & Shen, H. (2026). Beyond the single turn: Reframing refusals as dynamic experiences embedded in the context of mental health support interactions with LLMs. https://arxiv.org/html/2602.01694v1

Trivers, R. L. (1971). The evolution of reciprocal altruism. *The Quarterly Review of Biology, 46*(1), 35–37.

Vaccaro, I. (2025). Defining socioecological reciprocity: Intentionality, mutualism or collateral effect. *People and Nature, 7*(5), 1005–1010. https://doi.org/10.1002/pan3.10685

Van der Goot, M. J., & Etzrodt, K. (2023). Disentangling two fundamental paradigms in human-machine communication research: Media equation and media evocation. *Human-Machine Communication, 6*, 17–30. https://doi.org/10.30658/hmc.6.2

Van Wagtendonk, A. (2026, February 5). Wisconsin lawmakers explore age verification requirements on companionship chatbots. *Wisconsin Public Radio*. https://www.wpr.org/news/chatbots-age-verification-companionship-minors-wisconsin-assembly-chatgpt

Weijers, D., & Munn, N. (2025). AI companions: Assessing the future risks and benefits to wellbeing. In M. J. Dennis & P. Königs (Eds.), *The future of digital wellbeing*. Amsterdam University Press.

Weinstein, N., Itzchakov, G., & Maniaci, M. R. (2025). Exploring the connecting potential of AI: Integrating human interpersonal listening and parasocial support into human-computer interactions. *Computers & Human Behavior: Artificial Humans, 4*, 100149. https://doi.org/10.1016/j.chbah.2025.100149

Whatmore, S. (1999). Hybrid geographies. In D. Massey, J. Allen, & P. Sarre (Eds.), *Human geography today* (pp. 22–39). Polity Press.

Whatmore, S. (2017). Hybrid geographies: Rethinking the 'human' in human geography. In *Environment: Critical essays in human geography* (pp. 411–428). Routledge.

Zhang, R., & Xie, L. (2025). Companion or code? Navigating uncertainty in human-AI relationships. *SSRN Electronic Journal*. https://doi.org/10.2139/ssrn.5930192

Zhao, E. Y., & Bowman, N. D. (2026). Simulated intimacy, but real connection: Perceived reciprocity and romantic closeness make Otome fun to play. In *Proceedings of the Hawaii International Conference on Systems Science* (pp. 2891–2900). University of Hawaii at Manoa.